\documentclass[]{mn2e}
\usepackage{graphicx}
\usepackage{epsfig}
\usepackage{amsmath}
\newcommand       \Angstrom     {\,{\rm \AA}}

\newcommand       \K            {\,{\rm K}}

\newcommand     \gtsim  {\lower.5ex\hbox{$\buildrel > \over \sim$}}
\newcommand     \ltsim  {\lower.5ex\hbox{$\buildrel < \over \sim$}}
\newcommand     \simgt  {\lower.5ex\hbox{$\buildrel > \over \sim$}}
\newcommand     \simlt  {\lower.5ex\hbox{$\buildrel < \over \sim$}}

\newcommand       \mum          {\,{\rm \mu m}}

\newcommand	  \Teff	        {T_{\rm eff}}

\newcommand       \simali       {\sim\,}

\title[C$_{60}^{+}$ as the carrier of 
the $\lambda$9577$\Angstrom$
and $\lambda$9632$\Angstrom$
diffuse interstellar bands]{
C$_{60}$ cation as the carrier of 
the $\lambda$9577$\Angstrom$
and $\lambda$9632$\Angstrom$
diffuse interstellar bands:
Further support from 
the VLT/X-Shooter spectra
}
\author[Nie, Xiang \& Li]{T.P.~Nie$^{1,2}$,
  F.Y.~Xiang$^{1,2}$\thanks{fyxiang@xtu.edu.cn},
  and Aigen Li$^{2}$\thanks{lia@missouri.edu}\\
  $^1$Hunan Key Laboratory for Stellar
              and Interstellar Physics
              and School of Physics and Optoelectronics,
              Xiangtan University, Hunan 411105, China\\
  $^2$Department of Physics and Astronomy,
                  University of Missouri,
                  Columbia, MO 65211, USA\\
                  }

\begin{document}

\date{}
\pagerange{\pageref{firstpage}--\pageref{lastpage}} \pubyear{2021}

\maketitle

\label{firstpage}
\begin{abstract}
Ever since their first detection over 100 years ago,
the mysterious diffuse interstellar bands (DIBs),
a set of several hundred broad absorption features
seen against distant stars
in the optical and near infrared wavelength range, 
largely remain unidentified. 
The close match both in wavelengths
and in relative strengths recently found
between the experimental absorption spectra
of gas-phase buckminsterfullerene ions 
(C$_{60}^{+}$) and four DIBs
at $\lambda$9632$\Angstrom$,
$\lambda$9577$\Angstrom$,
$\lambda$9428$\Angstrom$ and
$\lambda$9365$\Angstrom$
(and, to a lesser degree, a weaker DIB at
$\lambda$9348$\Angstrom$)
suggests C$_{60}^{+}$ as a promising carrier for these DIBs.
However, arguments against the C$_{60}^{+}$ identification
remain and are mostly concerned with the large variation
in the intensity ratios of the $\lambda$9632$\Angstrom$
and $\lambda$9577$\Angstrom$ DIBs.
In this work
we search for these DIBs
in the ESO VLT/X-shooter archival data and identify
the $\lambda$9632$\Angstrom$,
$\lambda$9577$\Angstrom$,
$\lambda$9428$\Angstrom$,
and $\lambda$9365$\Angstrom$ DIBs
in a sample of 25 stars.
While the $\lambda$9428$\Angstrom$
and $\lambda$9365$\Angstrom$ DIBs
are too noisy to allow any reliable analysis,
the $\lambda$9632$\Angstrom$
and $\lambda$9577$\Angstrom$ DIBs
are unambiguously detected and,
after correcting for telluric
water vapor absorption, their correlation
can be used to probe their origin.
To this end, we select a subsample
of nine hot, O- or B0-type stars of which
the stellar Mg{\sc ii} contamination
to the $\lambda$9632$\Angstrom$ DIB
is negligibly small.
We find that their equivalent widths,
after normalized by reddening to eliminate
their common correlation with the density of
interstellar clouds, exhibit a tight, positive correlation,
supporting C$_{60}^{+}$ as the carrier
of the $\lambda$9632$\Angstrom$
and $\lambda$9577$\Angstrom$ DIBs.
\end{abstract}
\begin{keywords}
ISM: dust, extinction --- ISM: lines and bands
           --- ISM: molecules
\end{keywords}

\section{Introduction}\label{sec:intro}
The diffuse interstellar bands (DIBs) are a series of
over 600 broad absorption bands seen in the optical
and near infrared (IR) spectra of stars as the starlight
passes through the diffuse interstellar clouds
(Sarre 2006, Fan et al.\ 2019, 
McCabe 2019, Linnartz et al.\ 2020).
They were first observed in 1919
at 5780$\Angstrom$ and 5797$\Angstrom$
in the spectrum of the distant supergiant $\zeta$ Persei
by Mary Lea Heger (1922), a then graduate student
at Lick Observatory, and nearly two decades later,
the interstellar origin of these bands was established
by Merrill \& Wilson (1938), based on the strong correlation
between the band strength and the interstellar reddening.

Almost every article on DIBs started by stating that, 
``... since the discovery that DIBs are interstellar'', 
as D\'esert et al.\ (1995) put it,
``the nature of their carriers is still unknown''. 
This remained true until Campbell et al.\ (2015, 2016a,b), 
Walker et al.\ (2015, 2017) and Campbell \& Maier (2018)
for the first time measured the gas-phase spectrum
of buckminsterfullerene cation (C$_{60}^{+}$) 
and found that the spectral characteristics
(i.e., wavelengths and relative strengths)
of gas-phase C$_{60}^{+}$ are in agreement 
with four DIBs at $\lambda$9365.2$\Angstrom$,
$\lambda$9427.8$\Angstrom$,
$\lambda$9577.0$\Angstrom$,
and $\lambda$9632.1$\Angstrom$,
arguably as well as a weaker DIB
at $\lambda$9348.4$\Angstrom$.
Further support for this identification was provided 
by space observations obtained with
the {\it Hubble Space Telescope} (HST)
which were free of telluric absorption
contamination (Cordiner et al.\ 2019).

The original idea of C$_{60}^{+}$
being a possible DIB carrier dates
back to the poineering work
by Foing \& Ehrenfreund (1994),
who linked the neon/argon matrix spectra of
C$_{60}^{+}$ recorded by Fulara et al.\ (1993)
to two strong DIBs at $\lambda$9577$\Angstrom$
and $\lambda$9632$\Angstrom$. 
A direct comparison was not possible, though,
because the inert gas matrix environment would
broaden the band widths and shift the wavelengths.
More recently, the presence of C$_{60}^{+}$
in the interstellar medium (ISM) has been revealed
through the detections of the 6.4, 7.1, 8.2
and 10.5$\mum$ emission features of
C$_{60}^{+}$ in reflection nebulae, 
planetary nebulae and the Large and Small 
Magellanic Clouds (Bern\'e et al.\ 2013,
Strelnikov et al.\ 2015).
These detections may not be too surprising,
as its parent molecule C$_{60}$,
first experimentally synthesized
by Kroto et al.\ (1985), 
has also been detected in
various astrophysical environments 
through its characteristic IR emission features
at 7.0, 8.45, 17.3 and 18.9$\mum$
(Cami et al.\ 2010; Sellgren et al.\ 2010;
Garc{\'{\i}}a-Hern{\'a}ndez et al.\ 2010, 2011;
Zhang \& Kwok 2011).

If the $\lambda$9348$\Angstrom$,
$\lambda$9365$\Angstrom$,
$\lambda$9428$\Angstrom$,
$\lambda$9577$\Angstrom$,
and $\lambda$9632$\Angstrom$ DIBs indeed
share the same carrier and arise from C$_{60}^{+}$,
their strengths should correlate.
Although the assignment of these five DIBs
to C$_{60}^{+}$ has gained wide acceptance,
challenge has been continuously posed.
Particularly,
the intensity ratio of the two strongest bands of
C$_{60}^{+}$ at 9577 and 9632$\Angstrom$ has
became a topic of discussion.
While the laboratory spectrum of gas-phase
C$_{60}^{+}$ indicates an intensity ratio
of $\simali$0.84 for the 9632$\Angstrom$
band to the 9577$\Angstrom$ band
(Campbell \& Maier 2018),
Galazutdinov et al.\ (2017) found that,
after correcting for the Mg{\sc ii} stellar
contamination through model atmosphere calculations,
the ratio of the equivalent width of
the $\lambda$9632$\Angstrom$ DIB
to that of $\lambda$9577$\Angstrom$ DIB
is variable within a broad range.
%
%
Given that both the 9632$\Angstrom$ band
and the 9577$\Angstrom$ band of C$_{60}^{+}$
originate from electronic transitions starting
from the same level in the $^2{\rm Au}$ ground state
(but also see Lykhin et al.\ 2019, Hrodmarsson et al.\ 2020),
such a varying value is not a priori expected.
As a result, Galazutdinov et al.\ (2017) argued against
the assignment of the $\lambda$9577$\Angstrom$
and $\lambda$9632$\Angstrom$ DIBs to C$_{60}^{+}$
since their relative strengths are too poorly
correlated to be caused by a single source.
%
%
Using close spectral standards to correct for
the Mg{\sc ii} stellar contamination,
Walker et al.\ (2017) examined some of the same
spectra of Galazutdinov et al.\ (2017) and found that,
within the uncertainties, the $\lambda$9577$\Angstrom$
and $\lambda$9632$\Angstrom$ DIBs
are somewhat correlated. They further argued that
the use of close spectral standards is superior to
model atmosphere calculations in correcting for
contamination by the Mg{\sc ii} stellar lines.
More recently, Galazutdinov et al.\ (2021)
again reported a lack of correlation
between the equivalent widths
of the $\lambda$9577$\Angstrom$
and $\lambda$9632$\Angstrom$ DIB
for a sample of 43 stars.

To test the C$_{60}^{+}$ assignment of DIBs,
we search for these DIBs
in the ESO VLT/X-shooter archival spectra
and examine their interrelations.
We base on the ESO VLT/X-shooter archival data
and identify the $\lambda$9577$\Angstrom$
and $\lambda$9632$\Angstrom$ DIBs 
superimposed on the stellar spectra
of a number of stars.
We find that their strengths are 
well correlated and therefore provide
further support for the C$_{60}^{+}$ 
assignment of DIBs.
%
%
%
This paper is organized as follows.
We first briefly describe 
in \S\ref{sec:data}
the ESO VLT/X-shooter data set.
In \S\ref{sec:results} we report and discuss
the results on the DIB search and measure 
the DIB strengths, and also explore 
their correlations.
The major conclusions are summarized
in \S\ref{sec:summary}.

\begin{table*}
\footnotesize
\begin{center}
\caption[]{\footnotesize
           Stellar Parameters and the Equivalent Widths
           of the $\lambda$9577$\Angstrom$ DIB ($W_{9577}$)
           and $\lambda$9632$\Angstrom$ DIB ($W_{9632}$)
           for Our Sample of Nine Stars
           in the VLT/X-Shooter Archive
                 }
\label{tab:stars+dibs}
\begin{tabular}{lcccccccc}
\noalign{\smallskip} \hline \hline \noalign{\smallskip}	
Target Star	&	Spectral Type	& $T_{\rm eff}$
                &	$B^a$	
	        &	$V^a$
                &	$\left(B-V\right)_0^b$	
                &	$E(B-V)$
		&   $W_{9577}$
                &   $W_{9632}$      \\
      	        &
	        &	(K)
                &	(mag)
		&	(mag)
                &	(mag)
                &	(mag)
	        &   (m${\rm \AA}$)
	        &   (m${\rm \AA}$)\\
\noalign{\smallskip} \hline \noalign{\smallskip}
2MASS J17253421-3423116	&	O5.5IV$^c$	  & 40,000          &	13.53 & 11.82	 &	$-$0.29	&	2.00$^d$	    &	385.8$\pm$46.5  &	379.2$\pm$32.3  	\\
4U1907+09	            &	O8.5Iab$^c$	  & 33,000          & 19.41 & 16.35	 &	$-$0.27	&	3.33$^d$    &	349.2$\pm$36.4  &	278.8$\pm$7.4	    \\
Cl Pismis 24 17	        &	O3.5III$^c$	  & 44,000          & 13.33 & 11.84	 &	$-$0.26	&	1.75$^d$	&	314.1$\pm$8.6   &	384.7$\pm$13.8    \\
B111	                &	O4.5V$^e$     & 42,850$^e$      & $-$	& $-$    &	$-$	    &	1.35$^e$	&	384.1$\pm$52.8  & 392.0$\pm$18.7	\\
B150	                &	B0V$^f$	      & 30,000    & $-$	  & $-$      &	$-$	    &	1.32$^f$	&	497.8$\pm$41.6  &	492.6$\pm$29.0	\\
B164	                &	O6Vz$^e$	  &	39,100$^e$         & $-$	  & $-$      &	$-$	    &	1.76$^e$	&	557.5$\pm$7.3	  &	544.1$\pm$24.0	\\
B215                  	&	B0-B1V$^e$	  & 28,000$^e$         & $-$	  & $-$      &	$-$	    &	1.85$^e$	&	466.1$\pm$77.8  &	423.1$\pm$10.7	\\
B289	                &	O9.7V$^e$	  &	33,800$^e$         & $-$	  & $-$      &  $-$	    &	1.73$^e$	&	570.9$\pm$57.5  &	436.7$\pm$21.3	\\
B311	                &	O8.5Vz$^e$	  &	35,950$^e$         & $-$	  & $-$      &  $-$	    &	1.62$^e$    &	699.7$\pm$55.0  &	645.7$\pm$24.6	\\
\hline
\noalign{\smallskip} \noalign{\smallskip}
\end{tabular}
\begin{description}
\item[$^{a}$] $B$ and $V$ photometric magnitudes
                      taken from {\sf http://cdsportal.u-strasbg.fr/}.
\item[$^{b}$] Intrinsic colors $\left(B-V\right)_0$
                      taken from Wegner (2014).
\item[$^{c}$] Stellar spectral types adapted from
                      {\sf http://cdsportal.u-strasbg.fr/}.
\item[$^{d}$] Color excesses
                      $E(B-V)\equiv \left(B-V\right)-\left(B-V\right)_0$.
\item[$^{e}$] Ram\'irez-Tannus et al.\ (2018).
\item[$^{f}$] Nielbock et al.\ (2001).
\end{description}
\end{center}
\end{table*}

\begin{figure*}
\centerline{
\includegraphics[scale=0.2,clip]{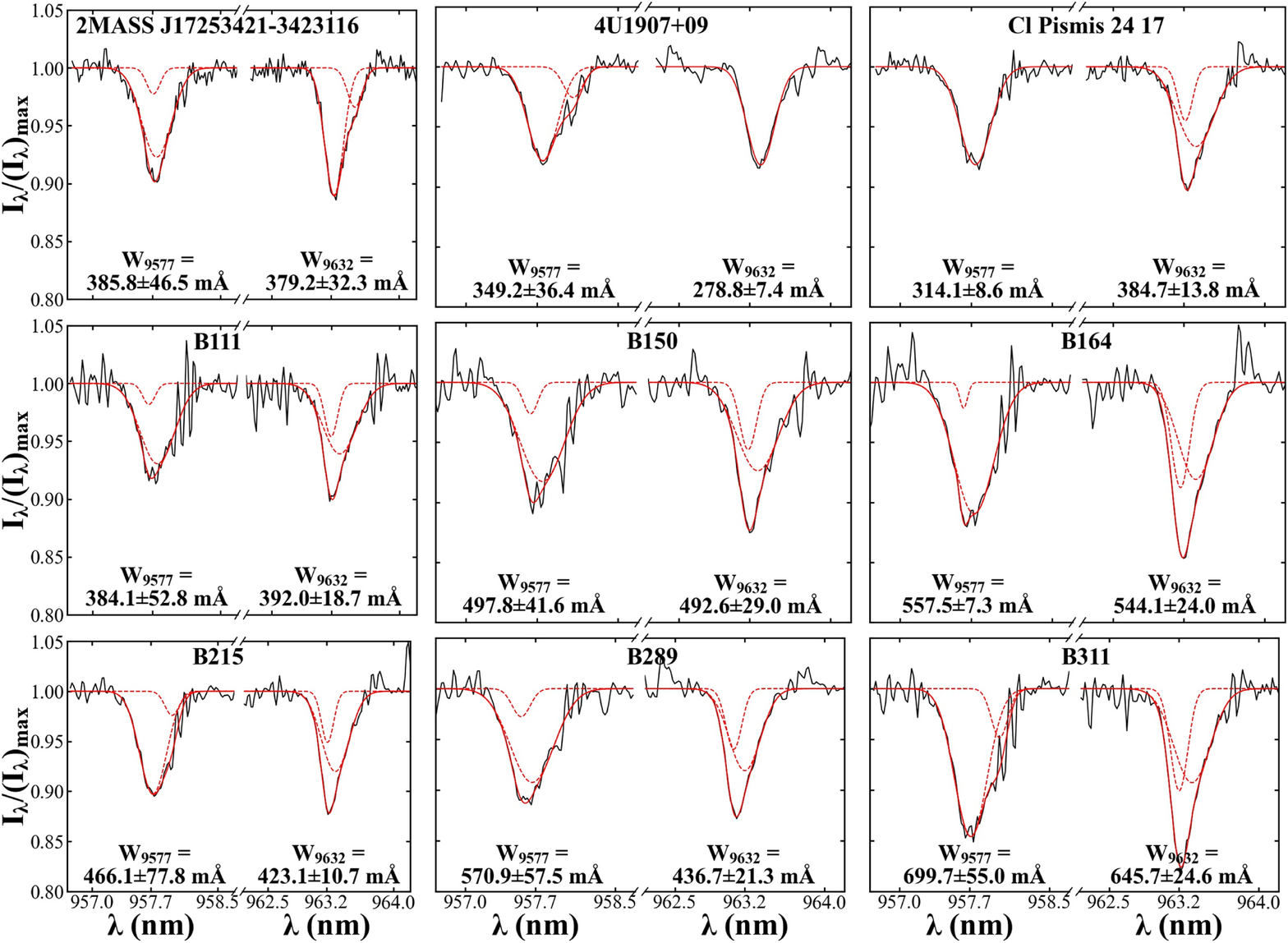}
}
\caption{
  \label{fig:dibs}
         The $\lambda$9577$\Angstrom$ and
         $\lambda$9632$\Angstrom$ DIBs
         (solid black lines) seen in the VLT/X-shooter
         spectra of nine target stars. Each DIB is fitted
         either by a single Gaussian profile
         or by a combination of two Gassuan profiles.
         The solid red lines show the fitted profiles,
         while the dotted red lines show the Gaussian
         fitting components.
         Labeled in each panel are the DIB equivalent widths.
         }
\end{figure*}

\section{ESO VLT/X-shooter Spectral Data Products}\label{sec:data}        
The X-shooter is the first of 
the 2nd generation instruments installed on 
the {\it European Southern Observatory} (ESO)
{\it Very Large Telescope} (VLT).
It is a very efficient single-target, 
medium-resolution spectrometer 
($R$\,$\simali$4,000--17,000),
covering the spectral range 
from 300 to 2,500\,nm
in a single exposure (Vernet et al.\ 2011).
We searched for the $\lambda$9348$\Angstrom$,
$\lambda$9365$\Angstrom$,
$\lambda$9428$\Angstrom$,
$\lambda$9577$\Angstrom$,
and $\lambda$9632$\Angstrom$ DIBs
from the X-shooter archival spectra
in the ESO Spectral Data Products.\footnote{%
  We note that in the literature these DIBs
  have been detected and their strengths
  have been reported for a number of sources.
  We prefer to search for these DIBs
  in archival data and derive their strengths
  by ourselves. This is because in the literature
  there is a certain arbitrarity in defining
  the continuum against the DIB absorption
  and the DIB strengths were often
  measured in different ways.
  When taking data from different sources,
  these differences actually play a role.
  }
%
%
We found 25 stars which simultaneously exhibit
the $\lambda$9365$\Angstrom$,
$\lambda$9428$\Angstrom$,
$\lambda$9577$\Angstrom$ and
$\lambda$9632$\Angstrom$ DIBs.
The $\lambda$9348$\Angstrom$ DIB,
the weakest band in the experimental
absorption spectrum of gas-phase C$_{60}^{+}$
(Campbell et al.\ 2015, Walker et al.\ 2017,
Campbell \& Maier 2018),
is not seen in the X-shooter spectra.
This does not necessarily mean that
the $\lambda$9348$\Angstrom$ DIB
is absent; instead, it may merely be weak
enough to have escaped detection.
%

It is interesting to note that
Galazutdinov et al.\ (2017) reported
the nondetection of the three weak absorption
bands of C$_{60}^{+}$ at 9348, 9365 and 9428$\Angstrom$
in a sample of 19 heavily reddened interstellar
sight lines observed from the ground at high
signal-to-noise (S/N).
The apparent absence of these weak bands
in sight lines where the $\lambda$9577$\Angstrom$
and $\lambda$9632$\Angstrom$ bands
are strong casts strong doubt
on the C$_{60}^{+}$ assignment.\footnote{%
    As noted by Linnartz et al.\ (2020),
    while Walker et al.\ (2015, 2016) reported 
    the detection of three weak DIBs at
    9348, 9365 and 9428$\Angstrom$
    which coincide with the weak absorption
    features seen in the gas-phase spectrum
    of C$_{60}^{+}$, these DIBs were not observed
    {\it simultaneously} along one line of sight,
    but merely complementary towards different targets.
    See Lallement et al.\ (2018) for
    an overview of the detectability of
    the DIBs attributed to C$_{60}^{+}$.
    }
However, the wavelength region
where the weak absorption bands of C$_{60}^{+}$
occur is heavily contaminated in ground-based
   studies due to strong telluric absorption.
   To circumvent the telluric contamination issues,
   Cordiner et al.\ (2019) obtained high S/N,
   telluric free HST spectra of seven heavily
   reddened stars and 
   reported unambiguous detections
   of two weak bands at 9365 and 9428$\Angstrom$.
   and one strong band at 9577$\Angstrom$.
   The intensity ratios of
   the $\lambda$9577$\Angstrom$,
   $\lambda$9428$\Angstrom$,
   and $\lambda$9365$\Angstrom$ DIBs 
   measured for early B stars were about 1.0:\,0.08:\,0.23,
   comparable to the experimental ratios
   of 1.\,0:\,0.15:\,0.25 derived from the laboratory
   spectrum of C$_{60}^{+}$ (see Cordiner et al.\ 2019).
   Unfortunately, the HST {\it  Space Telescope
   Imaging Spectrograph} (STIS) grating setting
   adopted by Cordiner et al.\ (2019) did not 
   allow them to survey
   the $\lambda$9632$\Angstrom$ DIB.
   Therefore, it is not possible to analyze 
   the relation between the $\lambda$9577$\Angstrom$
   and $\lambda$9632$\Angstrom$ DIBs
   based on the telluric free HST/STIS spectra.
%
%

\begin{figure*}
\centerline{
\includegraphics[scale=0.36,clip]{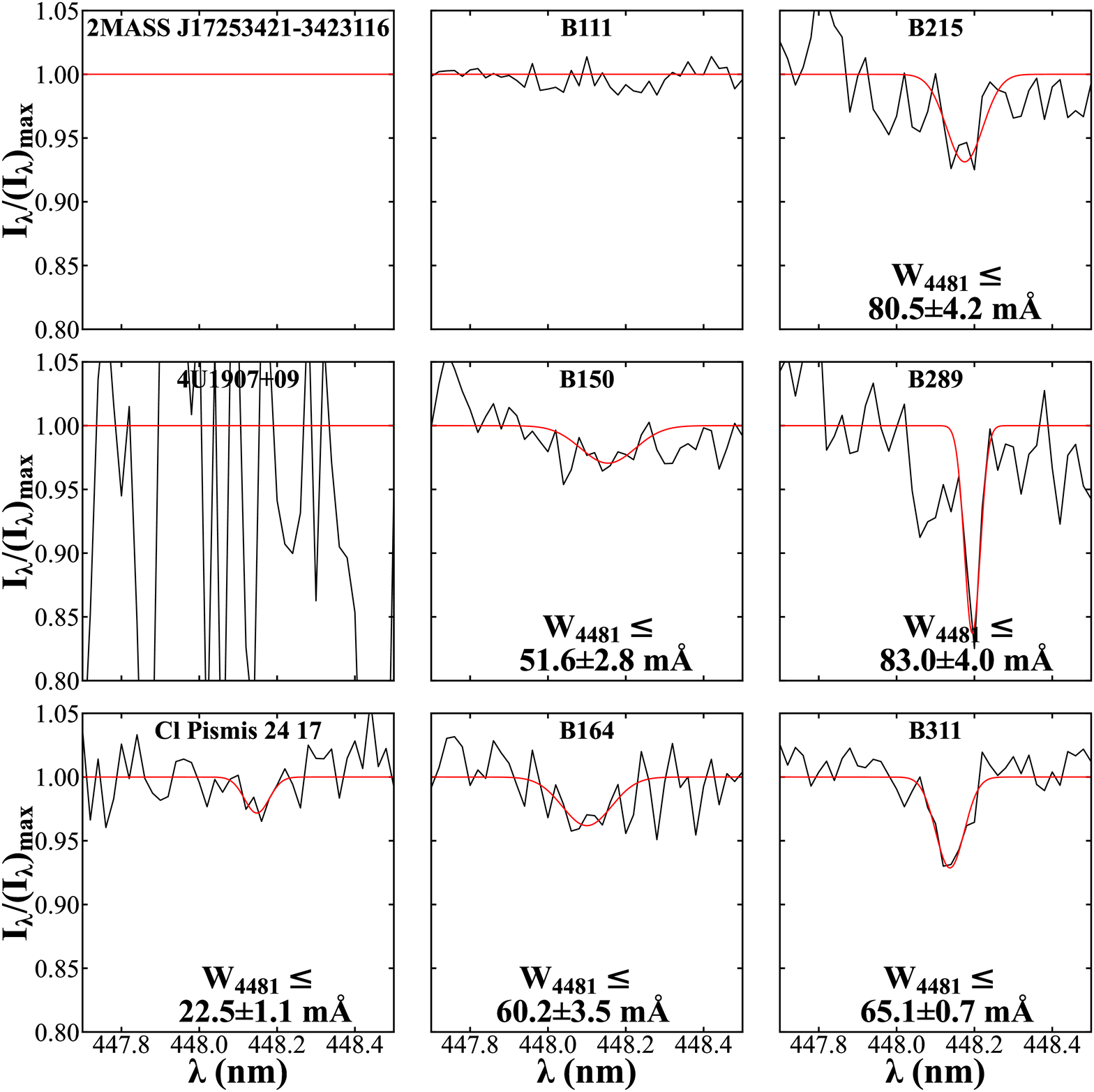}
}
\caption{
         \label{fig:MgII}
         The stellar Mg{\sc ii} line at 4481$\Angstrom$
         is absent or very weak in the X-shooter spectra
         (solid black line) of our target stars.
         For 2MASS J17253421-3423116,
         there is no X-shooter data.
         For 4U1907+09, the X-shooter spectrum
         is too noisy. 
         Labeled in each panel 
         is the equivalent width of
         the 4481$\Angstrom$ stellar Mg{\sc ii} line.   
         }
\end{figure*}

As mentioned earlier,
there are large numbers of telluric water vapor
absorption lines in the wavelength range of
C$_{60}^{+}$ bands.
We employ {\sf Molecfit} to correct X-shooter archive
data products for telluric absorption.
{\sf Molecfit} is a tool to correct for telluric absorption lines
based on synthetic modeling of the atmospheric transmission
which can be used with data obtained with various ground-based
telescopes and instruments
(Smette et al.\ 2015, Kausch el al.\ 2015).
Also, the $\lambda$9632$\Angstrom$ DIB
coincides and therefore often blends with the stellar
Mg{\sc ii} absorption lines at 9631.9 and 9632.4$\Angstrom$.
However, it is far from trivial to correct for contamination
by the stellar Mg{\sc ii} lines. 
Nevertheless, it is well recognized that while
the stellar Mg{\sc ii} lines are strong in late B stars,
they are negligibly weak in hot, O- and early B-type
stars of effective temperatures $\Teff\simgt20,000\K$
(e.g., see Figure~2 of Galazutdinov et al.\ 2017).
Therefore, we focus on a subsample of nine stars
which consists of seven O stars and two B0 stars
(see Table~\ref{tab:stars+dibs}).
In Figure~\ref{fig:dibs} we show the X-shooter spectra 
of the $\lambda$9577$\Angstrom$ and
$\lambda$9632$\Angstrom$ DIBs of these stars.
To verify that these stars are indeed weak in Mg{\sc ii}
absorption, we show in Figure~\ref{fig:MgII}
their Mg{\sc ii} stellar absorption line at 4481$\Angstrom$.
Except for 2MASS\,J17253421-3423116
for which there is no X-shooter data
around 4481$\Angstrom$,
the 4481$\Angstrom$ Mg{\sc ii} stellar line
is indeed either absent or very weak
in all the other eight stars.  
Therefore, the correction for 
stellar contamination by Mg{\sc ii}
to the $\lambda$9632$\Angstrom$ DIB
is unnecessary for our stars of which 
the effective temperatures all exceed 28,000$\K$.
%

\section{The $\lambda$9577$\Angstrom$
            and $\lambda$9632$\Angstrom$
            DIBs and Their Correlations}\label{sec:results}
As described in \S\ref{sec:data},
although the X-shooter spectra of our target stars
show evidence for the presence of all four DIBs
at $\lambda$9365$\Angstrom$,
$\lambda$9428$\Angstrom$,
$\lambda$9577$\Angstrom$
and $\lambda$9632$\Angstrom$ DIBs,
the spectral profiles of  the $\lambda$9365$\Angstrom$
and $\lambda$9428$\Angstrom$ DIBs are too noisy to
allow for any reliable quantitative analysis. 
In contrast, as shown in Figure~\ref{fig:dibs},
the $\lambda$9577$\Angstrom$
and $\lambda$9632$\Angstrom$ DIBs,
the strongest absorption features 
in the experimental spectrum of
gas-phase C$_{60}^{+}$ (Campbell et al.\ 2015,
Walker et al.\ 2017, Campbell \& Maier 2018),
are unambiguously detected in the X-shooter spectra.
As mentioned earlier, the stellar Mg{\sc ii}
absorption lines which would pollute
the $\lambda$9632$\Angstrom$ DIBs
are absent or very weak in hot O- and
early B-type stars. We therefore confine
ourselves to nine O and B0 stars along
the lines of sight to which the stellar Mg{\sc ii}
contamination is negligible. 
After correcting for the telluric contamination
with the {\sf Molecfit} tool, 
we fit the absorption profiles of
the $\lambda$9577$\Angstrom$
and $\lambda$9632$\Angstrom$ DIBs
by a single Gaussian profile or a combination
of two Gaussian profiles.\footnote{%
   We are only interested in the DIB band strength 
   (i.e., the area obtained by integerating
   the absorption band over wavelength)
   and thus the exact functional profile
   adopted to fit the band is not critical.
   We have actually also tried to fit the DIB 
   absorption bands in terms of Lorentzian 
   or Drude profiles which are expected 
   for damped harmonic oscillators (see Li 2009).
   However, both Lorentzian and Drude profiles
   are too broad in the blue and red wings to
   reproduce the DIB absorption bands.
   }
The $\lambda$9577$\Angstrom$
and $\lambda$9632$\Angstrom$ DIB profiles are often
asymmetrical, therefore a single Gaussian profile is often
not sufficient and two or more Gaussian profiles
are required to closely reproduce the observed DIB profiles
(e.g., see Rawlings et al.\ 2014).
Indeed, as shown in Figure~\ref{fig:dibs}, a combination
of two Gaussian profiles are needed for the majorities of
the target stars (8/9 for both DIBs).
In principle, we could also fit the DIB profiles
in terms of one or two Lorentzian functions.
This really does not matter since we are only
interested in the area integrated over the DIB
absorption profile. 

We determine the absorption strengths of
the $\lambda$9577$\Angstrom$
and $\lambda$9632$\Angstrom$ DIBs
from the fitted Gaussian profile(s) 
in terms of their equivalent widths,
$W_{9577}$ and $W_{9632}$.
The equivalent width of a DIB is a measure of
the DIB absorption strength which characterizes
the ``width'' of a ``virtual'' rectangle whose area
is equal to the area in the DIB profile and whose 
height is equal to the continuum level of the DIB profile.
We tabulate in Table~\ref{tab:stars+dibs} the measured
equivalent widths ($W_{9577}$, $W_{9632}$)
and their associated uncertainties of 
the $\lambda$9577$\Angstrom$
and $\lambda$9632$\Angstrom$ DIBs
for the nine target stars.
%

\begin{figure*}
\centerline{
\includegraphics[scale=0.39,clip]{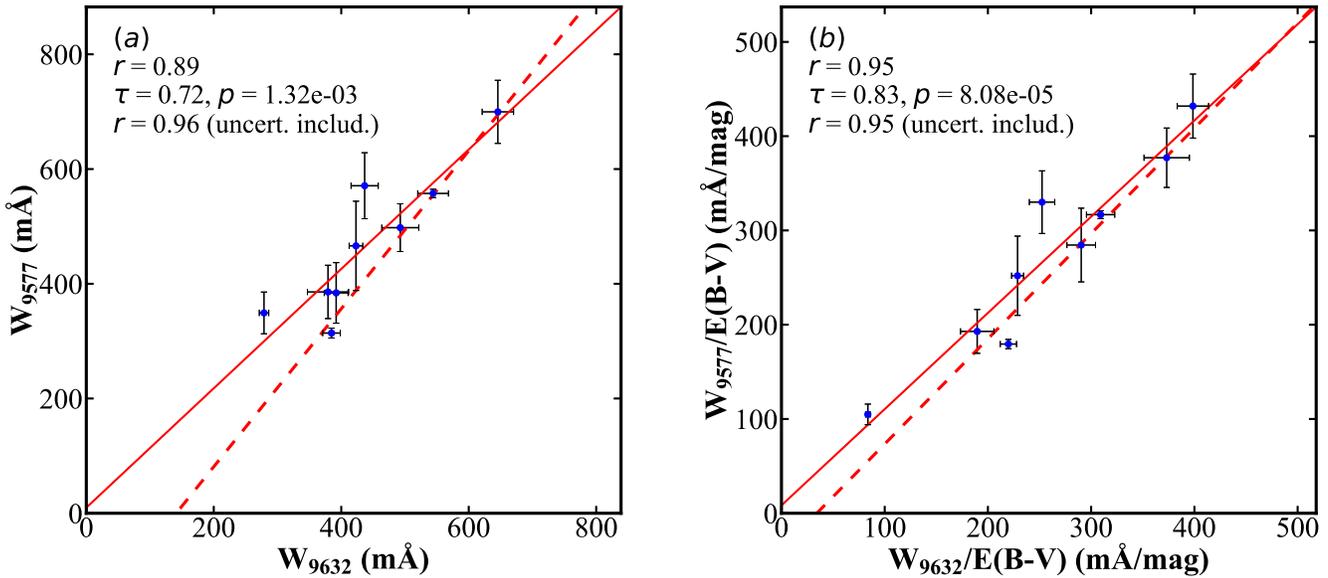}
}
\caption{
  \label{fig:9577vs9632}
         Left panel (a): Correlation diagram of the equivalent widths
         of the $\lambda$9577$\Angstrom$ DIB
         with that of the $\lambda$9632$\Angstrom$ DIB
         seen in our sample of nine target stars.
         Right panel (b): Correlation diagram of
         the {\it reddening-normalized} equivalent widths
         of the $\lambda$9577$\Angstrom$ DIB
         with that of the $\lambda$9632$\Angstrom$ DIB
         for our sample of nine stars.
         Labeled in both panels are
         the Pearson correlation coefficient $r$, 
         the Kendall's $\tau$ coefficient 
         and the significance level $p$.
         In each panel, the dashed line shows
         the correlation when the measurement
         uncertainties are taken into account,
         while the solid line shows the correlation 
         when we ignore the measurement uncertainties.
         }
\end{figure*}

If C$_{60}^{+}$ indeed causes
both the $\lambda$9577$\Angstrom$
and $\lambda$9632$\Angstrom$ DIBs,
their equivalent widths should correlate.
While a good correlation between two DIBs
does not necessarily mean that they must share
a common carrier, a non-correlation implies
that different carriers are involved
(e.g., see Moutou et al.\ 1999).
We conduct a correlation analysis on
$W_{9577}$ and $W_{9632}$
to investigate whether the $\lambda$9577$\Angstrom$
and $\lambda$9632$\Angstrom$ DIBs are related,
so as to determine whether they are from the same
carrier (e.g., C$_{60}^{+}$).
As illustrated in Figure~\ref{fig:9577vs9632}a,
with a Pearson correlation coefficient of $r\approx0.89$
and a Kendall correlation coefficient of $\tau\approx0.72$ 
at a significance level of $p\approx1.32\times10^{-3}$,
it is apparent that $W_{9577}$ and $W_{9632}$
are well correlated.
The correlation coefficient becomes 
$r\approx0.96$ when the measurement 
uncertainties are taken into account.

On average, the strength ratio
of the $\lambda$9632$\Angstrom$ DIB
to the $\lambda$9577$\Angstrom$ DIB
is $\simali$0.94, which agrees reasonably
well with that of the experimental spectrum
of gas-phase C$_{60}^{+}$
($\simali$0.84, Campbell \& Maier 2018).
%
%
The discrepancy of $\simali$12\% 
between the observed ratio of 
$W_{9632}/W_{9577}\approx0.94$
and the experimental ratio of
$W_{9632}/W_{9577}\approx0.84$
is acceptable.
We note that the reported experimental 
intensity ratio was not for pure C$_{60}^{+}$,
but actually for small He-tagged
C$_{60}^{+}$--He$_{\rm n}$ ($n$\,=\,1--3)
ion complexes in an ion trap 
(Campbell \& Maier 2018).
%
Upon tagged by He atoms, 
the laboratory rest wavelengths of C$_{60}^{+}$
are slightly redshifted. 
As the redshift is linearly dependent 
on the number of He atoms
(see Gatchell et al.\ 2019),
the absorption wavelengths of bare C$_{60}^{+}$
can be extrapolated from that of 
C$_{60}^{+}$--He$_{\rm n}$ complexes
(e.g., see Campbell et al.\ 2016b,
Spieler et al.\ 2017).\footnote{%
  The linear dependence of the wavelength
  redshift on the number of tagged He atoms
  remains valid up to 32 He atoms.
  From 33 He atoms onwards the redshift
  is not linear anymore (see Gatchell et al.\ 2019).
  }
The absence of actual data for 
bare C$_{60}^{+}$ still leaves 
a concern about the exact 
$W_{9632}/W_{9577}$ intensity ratio. 
Moreover, even the nature of the electronic 
transitions responsible for 
the 9632 and 9577$\Angstrom$ absorption bands 
seen in C$_{60}^{+}$ remains unknown 
(e.g., see Lykhin et al.\ 2019, 
Hrodmarsson et al.\ 2020).

Nevertheless, as mentioned earlier, 
such a correlation between $W_{9577}$ 
and $W_{9632}$ does not necessarily 
mean that the $\lambda$9577$\Angstrom$
and $\lambda$9632$\Angstrom$ DIBs must arise
from the same carrier. The equivalent widths of
two DIBs resulting from different carriers
for a random sample of lines of sight
may somewhat correlate {\it if} their carriers
are present in the interstellar medium (ISM)
of different lines of sight
in proportional abundances.
In general, most interstellar quantities
show a linear increase with the densities
of the interstellar clouds measured by gas,
e.g, the hydrogen column density $N_{\rm H}$,
or by dust, e.g., the interstellar reddening
$E(B-V)$ or extinction $A_V$
(e.g., see Witt et al.\ 1983;
Seab \& Snow 1984;
Fitzpatrick \& Massa 1988;
Moutou et al.\ 1999;
Xiang et al.\ 2011, 2017;
Li e al.\ 2019).
The correlation between two DIBs may
only indicate the coexistence of two separate
agents in the ISM, of which the abundances
both are correlated with $N_{\rm H}$ or $E(B-V)$.
Indeed, it is long recognized that the DIB strengths
are correlated with the inerstellar reddening $E(B-V)$.
To eliminate the common correlation with $E(B-V)$,
we therefore normalize the DIB equivalent widths
by $E(B-V)$ and examine the correlation between
$W_{9577}/E(B-V)$ and $W_{9632}/E(B-V)$.
As shown in Figure~\ref{fig:9577vs9632}b,
with a Pearson correlation coefficient of 
$r\approx0.95$\footnote{%
  The correlation coefficient essentially
  remains unchanged when the measurement 
  uncertainties are taken into account.
  }
and a Kendall correlation coefficient of $\tau\approx0.83$ 
at a significance level of $p\approx8.08\times10^{-5}$,
$W_{9577}/E(B-V)$ and $W_{9632}/E(B-V)$
are clearly well correlated.
%
%
This supports that a common carrier is at the origin
of the $\lambda$9577$\Angstrom$
and $\lambda$9632$\Angstrom$ DIBs.
We derive a mean ratio of
$W_{9632}/E(B-V):W_{9632}/E(B-V)\approx0.95$,
which also agrees reasonably well
with the experimental band ratio
of gas-phase C$_{60}^{+}$
($\simali$0.84, Campbell \& Maier 2018).

%

%
%

\begin{figure*}
\centerline{
\includegraphics[scale=0.39,clip]{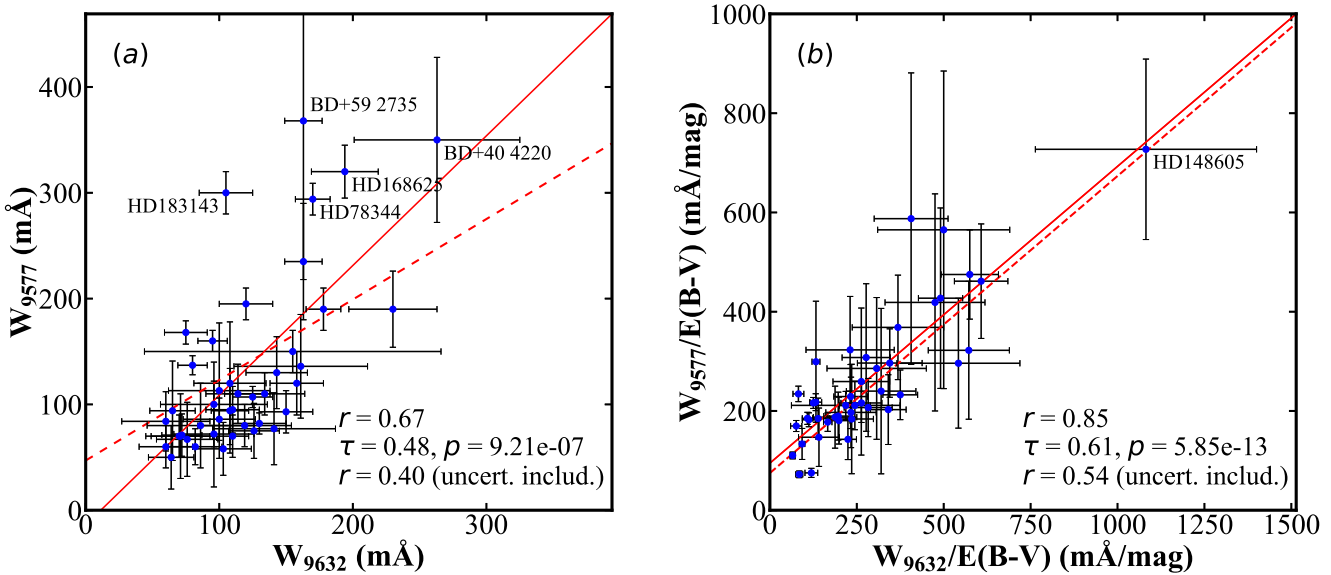}
}
\caption{
         \label{fig:9577vs9632Krelowski}
         Left panel (a):
         Correlation diagram of the equivalent widths
         of the $\lambda$9577$\Angstrom$ DIB
         with that of the $\lambda$9632$\Angstrom$ DIB
         for the sample of 43 stars
         of Galazutdinov et al.\ (2021)
         which are listed here
         in Table~\ref{tab:Krelowski2021}.
         The correlation coefficient
         (with the measurement uncertainties included)
         of $r\approx0.40$
         derived here closely agrees with that derived
         by Galazutdinov et al.\ (2021; $r\approx0.37$).         
         Right panel (b): Same as (a) but for
         the {\it reddening-normalized}
         equivalent widths. It is apparent that,
         upon normalization, the correlation
         appreciably improves.
         The correlation coefficients largely
         remain unchanged if HD\,148605,
         an outlier, is excluded.
         In each panel, the dashed line shows
         the correlation when the measurement
         uncertainties are considered,
         while the solid line shows the correlation 
         when the measurement uncertainties
         are ignored.
         }
\end{figure*}

\begin{table*}
\footnotesize
\begin{center}
\caption[]{\footnotesize
           Stellar Parameters and the Equivalent Widths
           of the $\lambda$9577$\Angstrom$ DIB ($W_{9577}$)
           and $\lambda$9632$\Angstrom$ DIB ($W_{9632}$)
           for the 43 Stars of 
	   Galazutdinov et al.\ (2021).
           }      
\label{tab:Krelowski2021}
\begin{tabular}{lccccccc}
\noalign{\smallskip} \hline \hline \noalign{\smallskip}	
Target Star	&	Spectral Type
                &	$B^a$	
	        &	$V^a$
                &	$\left(B-V\right)_0^b$	
                &	$E(B-V)^c$
		&   $W_{9577}$
                &   $W_{9632}$      \\
      	        &
	    &	(mag)
		&	(mag)
                &	(mag)
                &	(mag)
	        &   (m${\rm \AA}$)
	        &   (m${\rm \AA}$)\\
\noalign{\smallskip} \hline \noalign{\smallskip}
BD-14 5037	&   B1.5Ia	           & 9.72   & 8.41	 &	-0.17	&	1.48	&	125$\pm$15	&	107$\pm$10	\\
CD-32 4348	&	B3Ia	           & 10.05  & 9.19	 &	-0.13	&	0.99	&	75$\pm$16	&	168$\pm$11	\\
BD+40 4220	&	O6.5-7f+O5.5-6f	   & 10.79  & 9.185	 &	-0.29	&	1.90	&	263$\pm$62	&	350$\pm$78	\\
BD+59 2735	&	B0Ib	           & 10.89  & 9.88	 &	-0.22	&	1.23	&	163$\pm$14	&	368$\pm$150	\\
Cyg OB2 No.\,7	&	O3If			   & 12     & 10.55  &	-0.31	&	1.76	&	163$\pm$14	&	235$\pm$55	\\
HD23180	    &	B1III			   & 3.88   & 3.83	 &	-0.21	&	0.26	&	141$\pm$46	&	77$\pm$34	\\
HD27778	    &	B				   & 6.499  & 6.337  &	-0.11	&	0.27	&	64$\pm$17	&	50$\pm$30	\\
HD36861	    &	O8III			   & 3.48   & 3.47	 &	-0.26	&	0.27	&	71$\pm$22	&	70$\pm$40	\\
HD37022	    &	O7Vp			   & 5.15   & 5.13	 &	-0.29	&	0.31	&	72$\pm$15	&	71$\pm$20	\\
HD37041	    &	O9.5IVp			   & 6.3    & 6.39	 &	-0.31	&	0.22	&	108$\pm$14	&	94$\pm$40	\\
HD40111	    &	B0/1II/III		   & 4.76   & 4.82	 &	-0.22	&	0.16	&	76$\pm$23	&	67$\pm$35	\\
HD54662	    &	O6.5Vz(n)+O7.5Vz   & 6.24   & 6.21	 &	-0.29	&	0.32	&	60$\pm$20	&	60$\pm$20	\\
HD57061	    &	O9II			   & 4.25   & 4.4	 &	-0.31	&	0.16	&	65$\pm$17	&	94$\pm$47	\\
HD76341	    &	O9.2IV			   & 7.392  & 7.163	 &	-0.28	&	0.51	&	134$\pm$30	&	110$\pm$20	\\
HD78344	    &	O9.5/B0(Ib)		   & 10.12  & 9	     &	-0.24	&	1.36	&	170$\pm$13	&	294$\pm$15	\\
HD80077	    &	B2Ia+e		 	   & 10.29  & 9	     &	-0.16	&	1.45	&	95$\pm$11	&	160$\pm$10	\\
HD91824	    &	O7V((f))z		   & 8.11   & 8.14	 &	-0.29	&	0.26	&	60$\pm$33	&	84$\pm$28	\\
HD104705	&	B0III/IV		   & 9.56   & 9.11	 &	-0.23	&	0.68	&	96$\pm$40	&	100$\pm$40	\\
HD113904	&	WC5+B0III		   & 5.5    & 5.53	 &	-0.23	&	0.20	    &	100$\pm$38	&	113$\pm$64	\\
HD136239	&	B2Iae		 	   & 8.86   & 7.95	 &	-0.16	&	1.07	&	120$\pm$20	&	195$\pm$15	\\
HD144470	&	B1V				   & 3.92   & 3.97	 &	-0.23	&	0.18	&	103$\pm$21	&	58$\pm$25	\\
HD145502	&	B2V				   & 4.05   & 4	     &	-0.21	&	0.26	&	158$\pm$20	&	120$\pm$30	\\
HD147165	&	O9.5(V)+B7(V)	   & 3.02   & 2.89	 &	-0.27	&	0.4	    &	230$\pm$33	&	190$\pm$36	\\
HD147888	&	B3/4V			   & 7.05   & 6.74	 &	-0.18	&	0.49	&	110$\pm$12	&	70$\pm$20	\\
HD148379	&	B2Iab			   & 5.95   & 5.37	 &	-0.16	&	0.74	&	80$\pm$11	&	137$\pm$9	\\
HD148605	&	B3V				   & 4.72   & 4.79	 &	-0.18	&	0.11	&	119$\pm$35	&	80$\pm$20	\\
HD148937	&	O6f?p			   & 7.12   & 6.71	 &	-0.31	&	0.72	&	143$\pm$23	&	130$\pm$34	\\
HD149038	&	O9.7Iab			   & 4.99   & 4.94	 &	-0.24	&	0.29	&	100$\pm$27	&	86$\pm$20	\\
HD149757	&	O9.2IVnn		   & 2.58   & 2.56	 &	-0.28	&	0.30	    &	96$\pm$30	&	72$\pm$50	\\
HD150136	&	O3.5-4III(f*)+O6IV & 5.78   & 5.65	 &	-0.26	&	0.39	&	108$\pm$27	&	120$\pm$58	\\
HD151804	&	O8Iaf			   & 5.29   & 5.22	 &	-0.3	&	0.37	&	126$\pm$19	&	75$\pm$26	\\
HD152408	&	O8Iape			   & 5.92   & 5.77	 &	-0.3	&	0.45	&	110$\pm$20	&	95$\pm$22	\\
HD152424	&	OC9.2Ia			   & 6.69   & 6.27	 &	-0.27	&	0.69	&	161$\pm$50	&	136$\pm$49	\\
HD153919	&	O6Iafcp			   & 6.78   & 6.51	 &	-0.31	&	0.58	&	114$\pm$24	&	110$\pm$28	\\
HD155806	&	O7.5V((f))z(e)	   & 5.52   & 5.53	 &	-0.29	&	0.28	&	86$\pm$40	&	80$\pm$40	\\
HD167264	&	O9.7Iab			   & 5.42   & 5.37	 &	-0.24	&	0.29	&	82$\pm$20	&	60$\pm$17	\\
HD167971	&	O8Iaf(n)+O4/5	   & 8.27   & 7.5	 &	-0.3	&	1.07	&	178$\pm$13	&	190$\pm$20	\\
HD168625	&	B6Iap			   & 9.78   & 8.37	 &	-0.06	&	1.47	&	194$\pm$25	&	320$\pm$25	\\
HD169454	&	B1Ia			   & 7.61   & 6.71	 &	-0.19	&	1.09	&	130$\pm$20	&	82$\pm$10	\\
HD170740	&	B2/3I			   & 5.96   & 5.72	 &	-0.16	&	0.40	    &	150$\pm$20	&	93$\pm$20	\\
HD183143	&	B6Ia			   & 8.08   & 6.86	 &	-0.06	&	1.28	&	105$\pm$20	&	300$\pm$20	\\
HD184915	&	B0.5IIIn		   & 4.93   & 4.96	 &	-0.22	&	0.19	&	70$\pm$25	&	70$\pm$20	\\
HD190603	&	B1.5Ia			   & 6.19   & 5.65	 &	-0.17	&	0.71	&	155$\pm$111	&	150$\pm$20	\\
\hline
\noalign{\smallskip} \noalign{\smallskip}
\end{tabular}
\begin{description}
\item[$^{a}$] $B$ and $V$ photometric magnitudes
                      taken from {\sf http://cdsportal.u-strasbg.fr/}.
\item[$^{b}$] Intrinsic colors $\left(B-V\right)_0$
                      taken from Wegner (2014).

\item[$^{c}$] Color excesses
                      $E(B-V)\equiv \left(B-V\right)-\left(B-V\right)_0$.
\end{description}
\end{center}
\end{table*}

Finally, we note that 
Galazutdinov et al.\ (2021) compiled
the spectral data of 43 lines of sight
for the $\lambda$9577$\Angstrom$
and $\lambda$9632$\Angstrom$ DIBs
obtained with different instruments on board
different telescopes.
They found that, with a correlation coefficient
of $r\approx0.37$ 
(with the measurement uncertainties included),
the equivalent widths of 
the $\lambda$9577$\Angstrom$ DIB
are poorly correlated with that of
the $\lambda$9632$\Angstrom$ DIB.
We re-analyze the correlation 
and derive an uncertainty-included 
correlation coefficient of $r\approx0.40$\footnote{%
  The small difference in the correlation 
  coefficient $r$ between ours and that
  of Galazutdinov et al.\ (2021)
  probably arises from
  the fact that the sight lines we consider
  (see Table~\ref{tab:Krelowski2021})
  may not be exactly the same as those
  considered by Galazutdinov et al.\ (2021);
  indeed, it is not very clear which sight lines
  were examined in Galazutdinov et al.\ (2021).
  } 
($r\approx0.67$ if we ignore the measurement 
 uncertainties; 
 see Figure~\ref{fig:9577vs9632Krelowski}a).
However, as discussed earlier, the relation
between two DIBs could be affected by their
common dependence on $E(B-V)$.
To cancel out the common dependence
on $E(B-V)$, we normalize the equivalent
widths of both DIBs by $E(B-V)$ and re-examine
their correlation.
Figure~\ref{fig:9577vs9632Krelowski}b shows
the correlation between $W_{9577}/E(B-V)$
and $W_{9632}/E(B-V)$ for the 43 lines of sight
of Galazutdinov et al.\ (2021). 
With a Pearson correlation coefficient of $r\approx0.85$
($r\approx0.54$ if the measurement uncertainties
are taken into account)
and a Kendall correlation coefficient of $\tau\approx0.61$ 
at a significance level of $p\approx5.85\times10^{-13}$,
the correlation between $W_{9577}/E(B-V)$ and
$W_{9632}/E(B-V)$ is considerably tighter than 
that between $W_{9577}$ and $W_{9632}$.
Therefore, we believe that these two DIBs
of the sample of Galazutdinov et al.\ (2021)
are also correlated.
The mean ratio of
$W_{9632}/E(B-V):W_{9577}/E(B-V)\approx0.90$
agrees well with the experimental band ratio
of $\simali$0.84 of gas-phase C$_{60}^{+}$
(Campbell \& Maier 2018).






\section{Conclusion}\label{sec:summary}
We have searched for the DIBs
at $\lambda$9632$\Angstrom$,
$\lambda$9577$\Angstrom$,
$\lambda$9428$\Angstrom$,
$\lambda$9365$\Angstrom$
and $\lambda$9348$\Angstrom$
which are attributed to C$_{60}^{+}$
in the ESO VLT/X-shooter spectral data archive.
We have identified 25 stars along the lines
of sight to which all these DIBs
(except the $\lambda$9348$\Angstrom$ DIB,
the weakest absorption band 
in the experimental spectrum of
gas-phase C$_{60}^{+}$)
are seen in the X-shooter spectra.
To avoid the Mg{\sc ii} stellar contamination
to the $\lambda$9632$\Angstrom$ DIB,
we focus on a subsample of nine O and B0
stars in which the Mg{\sc ii} stellar lines
are absent or very weak.
It is found that, after normalized by
reddening (to eliminate their common correlation
with the density of interstellar clouds), the equivalent
widths of the $\lambda$9632$\Angstrom$
and $\lambda$9577$\Angstrom$ DIBs
are well correlated,
whereas the X-shooter spectra for
the $\lambda$9428$\Angstrom$
and $\lambda$9365$\Angstrom$ DIBs
are too noisy to allow any reliable quantitative analysis.
We conclude that the correlation
found between the strengths of
the $\lambda$9632$\Angstrom$
and $\lambda$9577$\Angstrom$ DIBs,
the strongest absorption bands 
in the experimental spectrum of
gas-phase C$_{60}^{+}$,
supports C$_{60}^{+}$
as the carrier of these DIBs. 

\section*{Acknowledgements}
We thank K.J.~Li, A.N.~Witt, X.J.~Yang
and the anonymous referee
for helpful suggestions and comments.
TPN and FYX are supported in part by 
the Joint Research Funds in Astronomy U2031114
under cooperative agreement between 
the National Natural Science Foundation of China 
and Chinese Academy of Sciences
and the CCST Milky Way Survey Dust and Extinction Project.
AL is supported in part by NSF AST-1816411.

\section*{Data Availability}
The data underlying this article will be shared 
on reasonable request to the corresponding authors.


\bsp
\label{lastpage}
\end{document}